\documentclass[aps,reprint,twocolumn,superscriptaddress]{revtex4-2}

\usepackage{amsmath,mathtools,amssymb,amsthm}
\usepackage{bm,upgreek} 
\usepackage{xcolor}
\usepackage{braket}
\usepackage{epstopdf}
\usepackage{graphicx}
\usepackage[colorlinks,linkcolor=blue,anchorcolor=blue,citecolor=blue,urlcolor=blue]{hyperref}

\newcommand{\nn}{{\nonumber}}

\newcommand{\up}{\uparrow}
\newcommand{\dn}{\downarrow}

\newcommand{\0}[1]{\mathbf{#1}}
\newcommand{\av}[1]{\left\langle #1 \right\rangle}

\newcommand{\w}{\omega}

\begin{document}
\title{Genuine pair density  wave order on the kagome lattice}

\author{Han-Yang Liu}
\author{Da Wang} \email{dawang@nju.edu.cn}
\affiliation{National Laboratory of Solid State Microstructures $\&$ School of Physics, Collaborative Innovation Center of Advanced Microstructures, Nanjing University, Nanjing 210093, China}

\author{Ziqiang Wang} \email{wangzi@bc.edu}
\affiliation{Department of Physics, Boston College, Chestnut Hill Massachusetts 02467, USA}

\author{Qiang-Hua Wang} \email{qhwang@nju.edu.cn}
\affiliation{National Laboratory of Solid State Microstructures $\&$ School of Physics, Collaborative Innovation Center of Advanced Microstructures, Nanjing University, Nanjing 210093, China}

\begin{abstract}
The pair density wave (PDW) is a novel superconducting state with non-zero center-of-mass momentum Cooper pairing in the absence of external magnetic fields.
Its realization in microscopic models as the ground state is very rare and extremely challenging, because a genuine PDW state is free of a uniform component or modulations by a pre-existing spin/charge density wave order at the same wavevector.
Here, we report the discovery of a genuine primary PDW phase in a two-orbital Hubbard model on the kagome lattice by state-of-art functional renormalization group studies.
It emerges out of competing orders over a wide physical parameter range suitable for realistic material realizations. The key ingredients in favor of the PDW order are the strongly sublattice and orbital polarized Bloch states on multiple Fermi pockets.
They force the zero-momentum Cooper pairing to involve the same sublattices and be suppressed by onsite Coulomb repulsion, while pairing between different sublattices to be dominated by different Fermi pockets with nonzero total momentum.
The degenerate PDW states at three momenta ${\bf M}_{1,2,3}$ on the Brillouin zone boundary exhibit novel intertwined order and can linearly combine into topologically nontrivial chiral PDW states. We propose that the model can be realized in multiorbital kagome materials such as CsCr$_3$Sb$_5$ as well as cold atom systems.
\end{abstract}
\maketitle

\emph{Introduction}. The pair density wave (PDW) is a macroscopic phase coherent superconducting (SC) state where the Cooper pairs condense with nonzero center-of-mass momentum (CMM). It is a paradigm shift of the Fulde–Ferrell–Larkin–Ovchinnikov (FFLO) states \cite{FF1964,LO1964} induced by a Zeeman splitting into a spontaneous SC quantum matter in the absence of external magnetic field \cite{Chen2004}. The novelty lies in that a genuine PDW state is a primary order with a composite SC order parameter that spontaneously breaks translation symmetry and oscillates spatially with zero average. This extraordinary SC state produces rich and profound symmetry enabled intertwined electronic states and vestigial order through staged melting of the composite order parameter, such as vestigial charge density wave (CDW), spin density wave (SDW) and pseudogap behaviors, and the cluster pairing of
multiple Cooper pairs and fractional SC flux quantization \cite{Berg_NP_2009,Fradkin2015,Agterberg2020}.

The interest in searching for this intriguing quantum state has intensified recently \cite{Chen2023}. In contrast to the very long wavelength oscillations of the FFLO state due to the small Zeeman splitting, the PDW oscillations driven by correlations in short coherence length superconductors can form over the atomic length scale and amenable to direct visualization by scanning tunneling microscopy \cite{Wang2026}.
Indeed, growing evidence for spatially periodic modulations of superconductivity in several materials platforms has emerged.
Nevertheless, it is important to separate an intrinsic spontaneous PDW order from a secondary PDW with spatial modulations at the same wavevector of a primary CDW order already present in the normal state.
In this case, the CMM of the Cooper pairs is the same as the reciprocal vector of the CDW lattice structure so that the modulations are within the CDW unit cell. The secondary PDW have been observed in
cuprates \cite{Shi2020,Lozano2022,Chen2025},
transition metal dichalcogenides \cite{Liu2021,Cao2024},
kagome metals \cite{Li2023,Deng2024,Yan2024},
iron-based superconductors \cite{Zhao2023},
and heavy fermion superconductors \cite{Gu2023}.
Although, these SC modulations are interesting in their own rights, we will not be concerned with the secondary PDW here. Evidence for spontaneous SC modulations at wavevectors that only emerge in the SC state are rare, but some evidence have been observed in
cuprates \cite{Du2020,Chen2022} near impurity \cite{Hamidian2016,Wang2025} or vortex \cite{Edkins2019},
confined iron-based superconductors \cite{Liu2023} and $^3$He superfluid \cite{Shook2020}.
In bulk systems, the most promising candidate for a primary PDW so far comes from kagome metals \cite{Chen2021,Han2025}.
Nevertheless, the modulating amplitudes are typically quite small compared to the coexisting uniform SC condensate. Whether the primary PDW state has been observed remains an open and debated issue \cite{Gao2024,Yin2024}.

On the theoretical side, although proposed in many previous studies
\cite{Berg2010,Jaefari2012,Pepin2014,Wang2015,Xu2019,Zhou2022,Jin2022,Coleman2022,Wu2023,Wu2023a,Wu2023c,Jiang2023,Jiang2023a,Shaffer2023,Castro2023,Tsvelik2023,Jiang2024,Ticea2024,Yao2025,Wang2025a,Panigrahi2025,Zhang2025,Wang_A_2025}, the existence of the primary PDW has not been convincingly obtained in concrete and realistic models.
The difficulty is natural. The usual zero-CMM Cooper pairing of time-reversed electronic states has a divergent susceptibility causing an instability of the Fermi surface for infinitesimal attraction. In contrast, at a nonzero CMM, the two electrons forming the Cooper pair are in general not occupying degenerate energy states. As a result, the PDW susceptibility is no longer infrared divergent, and is thus generically disfavored over the zero-CMM pairing. The \emph{first} challenge for the primary PDW formation under a finite pairing attraction is to frustrate the formation of uniform zero-CMM Cooper pairs, opening the possibility for pairing at nonzero CMM $\0Q_p$. In order to achieve a primary PDW at $\0Q_p$, however, there is a \emph{second} challenge. The interactions in the particle-hole (PH) channel should at most produce a subleading CDW or SDW at momentum $\0Q_p$ compared to the leading PDW order at $\0Q_p$ in the Cooper channel. In the opposite case, the PDW becomes secondary and only describes SC modulations within the unit cell of the density wave order.

In this work, we introduce a two-orbital Hubbard model on the kagome lattice and show that the combined sublattice and multiorbital polarization and electronic correlation overcomes both the challenges and produce a genuine primary PDW as the ground state. Using the unbiased FRG approach, the PDW order is observed to emerge out of spin-fluctuation mediated pairing as the primary electronic order under sizable multiorbital Fermi pockets away from the vHS and under very generic local repulsive interactions. The PDW has three degenerate components at momenta ${\bf Q}_p={\bf M}_{1,2,3}$ on the Brillouin zone boundary. The FRG reveals a novel intertwined electronic order that breaks time-reversal symmetry with charge loop-current, and the Josephson coupling among the three PDW components can form a chiral topological PDW state.

\begin{figure}
\centering
\includegraphics[width=0.9\linewidth]{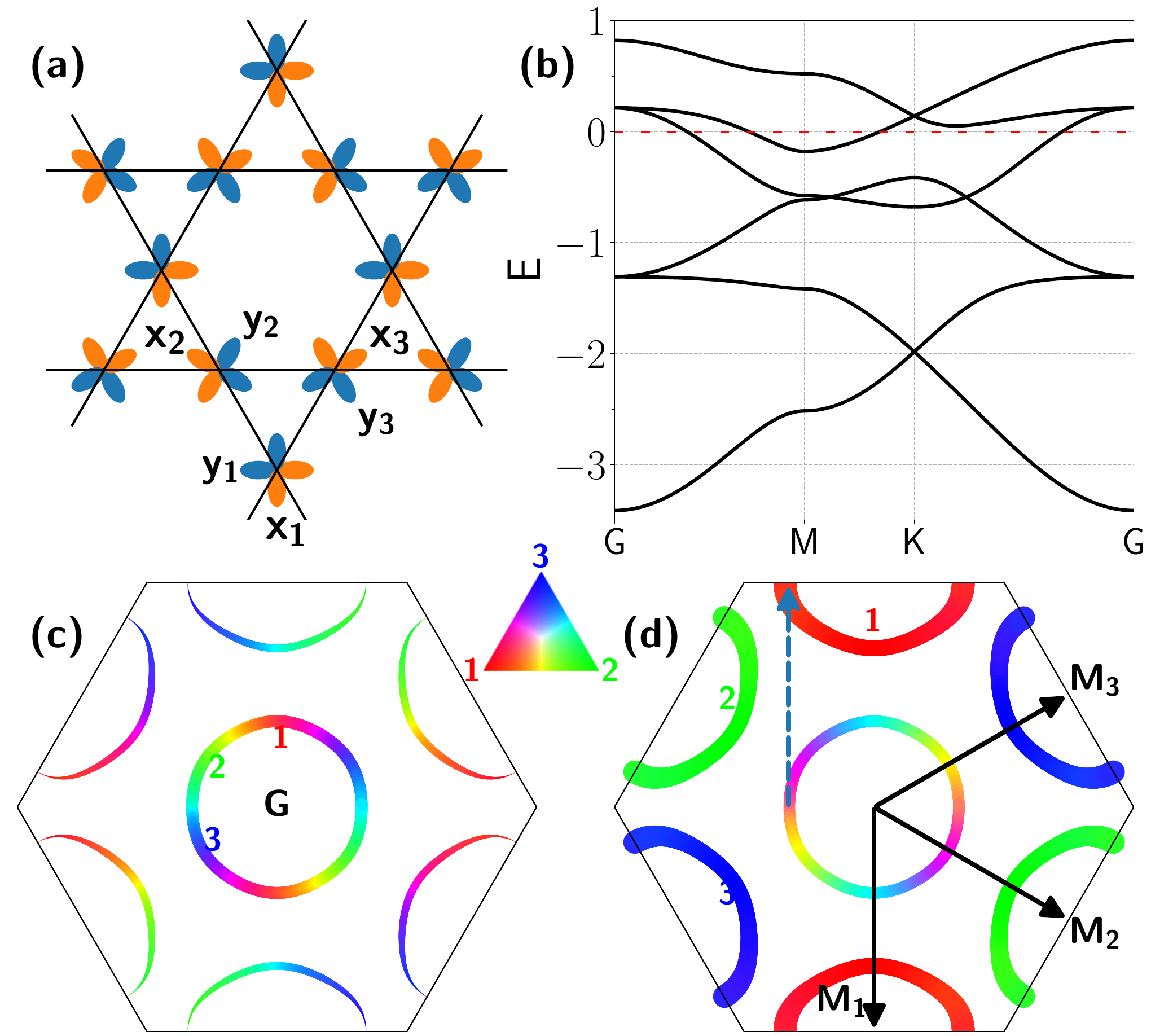}
\caption{(a) A sketch of the kagome lattice with three sublattices ($1,2,3$) and two local $x$ and $y$ orbitals. (b) Band dispersion along a high symmetry path. (c) Color-scaled sublattice content of the $x$ orbital on Fermi surfaces. The line width encodes the $x$ orbital content. (d) Same as (c) but for the $y$ orbital.
The solid arrows indicate the three M vectors associated with the pockets. The dashed arrow shows $\0G$-$\0M$ scattering with similar sublattice content.}
\label{fig:model}
\end{figure}

\emph{Model and Method}. The model we consider is illustrated in Fig.~\ref{fig:model}(a). There are two orbitals on each site, which transform as $X$ and $Y$ (or $XZ$ and $YZ$) globally under planar rotations and reflections. By recombining the global orbitals, we adopt the local orbital basis where $x$-orbital points to the center of the triangles and $y$-orbital is orthogonal to $x$. This makes it possible for the site local crystal-field energies to be diagonal. The tight-binding part of the Hamiltonian can be written as
\begin{align}
H_0=\sum_{\av{ij}abs} (t_{ij}^{ab} c_{ias}^\dag c_{jbs} + h.c.) + \sum_{ias}(E_a-\mu) n_{ias},
\end{align}
where $a/b$ stands for the (henceforth) local $x$ and $y$ orbitals, $s$ denotes spin, $\av{ij}$ denotes the nearest neighbor (NN) bonds, $E_a$ is the onsite energy of the $a$-orbital, and $\mu$ is the chemical potential. The electron creation operator $c_{ias}^\dag$ and particle number operator $n_{ias}=c_{ias}^\dag c_{ias}$ are standard. The hopping integrals between the $a$ and $b$ obey the Slater-Koster rule, $t_{ij}^{ab}=\cos\theta_a \cos\theta_b t_{pp\sigma} + \sin\theta_a\sin\theta_b t_{pp\pi}$, where $\theta_{a/b}$ is the angle between the $a/b$ orbital orientation and the bond direction, and $t_{pp\sigma}$ ($t_{pp\pi})$ is the hopping integral for orbitals projected along (orthogonal) to the bond direction. For illustrative purposes, we set $t_{pp\sigma}=5/8$, $t_{pp\pi}=-1/8$, $E_x=-0.055$, $E_y=1.182$. The chemical potential is set as $\mu=1.36$ unless specified otherwise, and all energies are assumed dimensionless. These parameters are tuned to mimic a realistic Fermi surface topology, which can appear in CsCr$_3$Sb$_5$ \cite{CsCr3Sb5}, for example. We also remark that our results hold equally when the two orbitals are switched by exchanging the site energies, and exchanging the Koster-Slater coefficients followed by a minus sign.

The band dispersion is presented in Fig.~\ref{fig:model}(b). The lower and upper three bands are mostly from $x$ and $y$ orbitals, respectively. When the inter-orbital hybridization is ignored, the lower and upper three bands would be two copies of the single-orbital kagome model, with the third and forth bands being the flat bands. The orbital hybridization reshapes the Dirac bands crossing each other at $\0K$ point, and distorts the flat bands, making them dispersive. Moreover, the upper band complex is narrower than the lower one in our choice of the Koster-Slater coefficients. The Fermi level ($E=0$) is chosen to cross the narrower complex to enhance correlation effects to be addressed.

We can see a hole-like pocket around $\0G$ coming from the distorted flat band, and electron-like pockets around $\0M_i$ on the zone boundary. Both types of Fermi pockets are parabolic around the respective centers, hence are not subject to vHS.  Fig.~\ref{fig:model}(c) shows the sublattice contents for the $x$ orbital on the Fermi surfaces. The three sublattice contents are clearly separated on the central $\0G$-pocket (denoted by the R-B-G colors and labeled by numbers), whereas they are pairwise mixed on the $\0M$-pockets on the zone boundary along the $\0G-\0M$ directions.
In comparison, Fig.~\ref{fig:model}(d) shows the sublattice contents on the Fermi surfaces associated with the $y$ orbitals. The thicker lines represent the larger spectral weight of the $y$ orbitals on the Fermi surfaces. The three sublattice contents are clearly separated on the $\0M$-pockets. On the $\0G$-pocket, the sublattice content is also separated, although to a lesser extent than on the $\0M$-pockets. The dashed arrow in Fig.~\ref{fig:model}(d) shows the part of the $\0G$-pocket dominated by the sublattice-1 (red), as compared to the almost pure sublattice-1 content on the $\0M_1$-pocket.

The electron-electron interactions are described by the standard two-orbital Hubbard model,
\begin{align}
H_I=&\sum_{ia}Un_{ia\up}n_{ia\dn}+\sum_{i,a\ne b}J_P c_{ia\up}^\dag c_{ia\dn}^\dag c_{ib\dn} c_{ib\up} \nn\\
&+\sum_{i,a<b,\sigma\sigma'}U'n_{ia\sigma}n_{ib\sigma'}+J_Hc_{ia\sigma}^\dag c_{ib\sigma} c_{ib\sigma'}^\dag  c_{ia\sigma'} ,
\end{align}
where $U$ and $U'$ are intra- and inter-orbital Coulomb repulsions, $J_H$ is the Hund's coupling, and $J_P$ is the pair hopping interaction. These four interactions are assumed to respect the standard relations: $U=U'+2J_H$ and $J_H=J_P$. The Hubbard $U$ is expected to suppress the onsite pairing, leaving room for intersite pairing on bonds.

The interactions can lead to intriguing correlation effects in the spin, charge and pairing channels. To treat all channels on equal footing and to take care of the orbital-sublattice polarization exactly, we resort to the state-of-art singular-mode FRG (SM-FRG) \cite{Wang2012,Wang2013,Tang2019,Yang2023}, see Sec.I of the Supplemental Materials (SM) \cite{sm} for self-contained details. During the FRG flow, we monitor the effective interactions $V_{\rm eff}$ in SC, SDW and CDW channels, versus a decreasing infrared cutoff energy scale $\Lambda$. The first divergent channel indicates the corresponding instability. The divergent energy scale $\Lambda_c$ represents the critical temperature $T_c$, and the corresponding eigen mode of the effective interaction (taken as scattering matrix between fermion bilinears) provides the structure of the operator for emerging order, such as the pairing function, the spin or charge structure.

\emph{Results and discussions}. We first discuss how our model meets the challenges for PDW. Figs.~\ref{fig:model}(c-d) clearly reveals that despite being away from the vHS, the model contains well-defined sublattice selectivity. The time-reversed pairs of the quasiparticle states on the Fermi surfaces, e.g. $\ket{\0k\up}$ and $\ket{-\0k\dn}$, occupy predominately the same sublattice and are thus unfavorable for forming zero CMM Cooper pairs in the presence of strong onsite Coulomb repulsion. The three $\0M$-vectors pointing to the zone boundary are marked in Fig.~\ref{fig:model}(d). Note that the sum of any two of the three M-vectors is equivalent to the third, modulo the reciprocal lattice momentum, and $\0M_i\sim -\0M_i$ since $2\0M_i$ is a reciprocal lattice momentum. Obviously, all inter-pocket pairings result in a PDW carrying a CMM close to one of the $\0M_i$-vectors. We overcome the first challenge to realize a primary PDW.

The Fermi surface topology might imply possible nesting effect. In the PH channel, nesting requires $\varepsilon_\0k\approx-\varepsilon_{\0k+\0Q}$ for $\0k$ on one pocket and $\0k+\0Q$ on the other. Thus the PH scattering between all electron-like $\0M$-pockets is not quasi-nested. The PH scattering between the hole-like $\0G$-pocket and the electron-like $\0M$-pockets is quasi-nested, but it is significantly weakened because of the sublattice selectivity, at least for intra-sublattice scattering. On the other hand, in the particle-particle (PP) channel, nesting requires $\varepsilon_\0k\approx\varepsilon_{-\0k+\0Q}$. So the $\0M$-pockets are mutually quasi-nested and involves different sublattices, whereas the hole-like $\0G$-pocket and electron-like $\0M$-pockets are not quasi-nested. Taken together, the two-orbital model materialize a situation where the PH scattering is not expected to be particularly strong, and may be even weaker than the PP scattering at momentum $\0M$. This is the feature we are after to overcome the second challenge to realize a primary PDW.

\begin{figure}
\includegraphics[width=\linewidth]{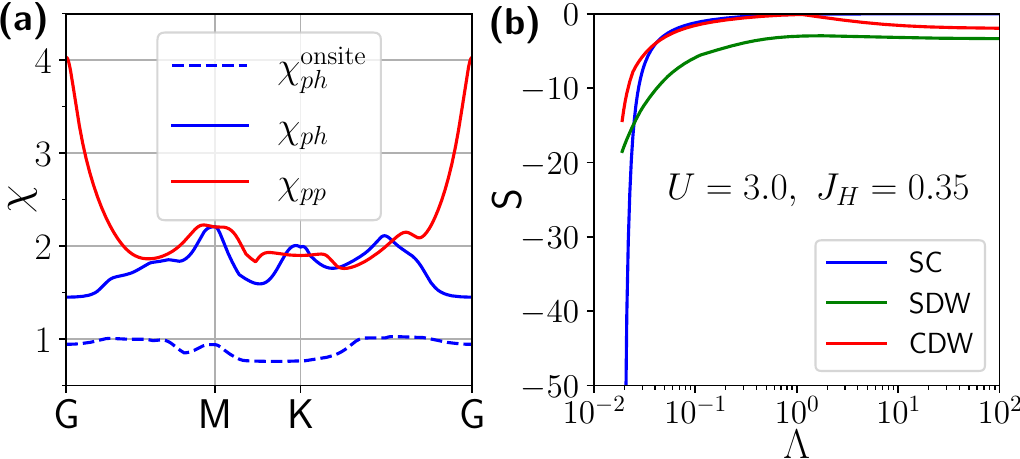}
\caption{(a) Bare susceptibilities in the PH (solid blue) and PP (red) channels. The dashed blue line presents the susceptibility with only onsite PH pairs. (b) FRG flow of the most negative eigenvalues of the effective interactions in the SC (blue), SDW (green), and CDW (red) channels.}
\label{fig:flow}
\end{figure}

To gain further insights, we plot the bare susceptibilities $\chi_{ph,pp}$ in the PH and PP channels in Fig.~\ref{fig:flow}(a). They are calculated from the leading eigenvalues of the susceptibility matrices in the fermion bilinear basis. The dashed blue line depicts the contribution to $\chi_{ph}(\0q)$ from only the propagation of onsite PH pairs,
which is favored by the bare local interactions. We see broad peaks between $\0G$ and $\0K$, slightly lower peaks between $\0G$ and $\0M$, and an even lower peak at $\0M$. The former two sets of peaks at small momenta clearly come from intra-pocket scattering, and the last set at the large momentum $\0M$ comes from inter-pocket scattering. The scattering between the hole-like $\0G$-pocket and the electron-like $\0M$-pocket would have been strong, but because of sublattice selectivity, it is weakened. The overall scale of this part in $\chi_{ph}$ is moderate and clearly free from divergences. The solid blue line is the calculated $\chi_{ph}$ when both onsite and bond PH pairs are included.
Comparing to the dashed blue line, the intersite bond contributions are quite significant. The enhencement must have come from bond-wise PH's. Now the susceptibility has the highest peak at $\0M$, the second peak near $\0K/2$, and the third peak near $\0K$. These new or enhanced peaks are caused by the intricate orbital and sublattice polarization on the multiple Fermi pockets.

The Cooper channel susceptibility $\chi_{pp}$ is calculated including the propagation of both onsite and on-bond pairs, which is shown by the solid red line. Clearly, $\chi_{pp}$ is the strongest at $\0G$, which is associated with the standard Cooper instability. We checked the corresponding eigen mode and found that it contains mainly onsite PP pairs, which will be suppressed by local repulsive interactions. There is another weaker but broader peak near $\0M$. The eigen mode there contains electron pairs on NN bonds connecting unequal sublattices. In momentum space, these components connect electrons on two different $\0M$ pockets. We note that the same eigen mode also contains some smaller contributions from onsite fermion pairs. They may be suppressed by local repulsive interactions, but may also survive in the presence of Hund's coupling, see Sec.II of SM \cite{sm} for illustration. Finally, we observe that the broad feature of $\chi_{pp}$ near $\0M$ is larger than the overall scale of $\chi_{ph}$ for local PH's (dashed blue line) and comparible to $\chi_{ph}$ enhanced by bond-wise PH's (solid blue line). This leaves the possibility that the PP channel would win over the PH one in the presence of interactions, which we address below.

We now discuss the correlation effects from SM-FRG calculations. For a typical interaction parameter $(U, J_H)=(3, 0.35)$, Fig.~\ref{fig:flow}(b) shows the leading negative eigenvalue, $S$, of the effective interaction $V_{\rm eff}(\0q)$ in the three channels (SC, SDW, CDW), versus the running energy scale $\Lambda$. At high energy scales, the SDW channel dominates, since the bare local interactions are repulsive. The CDW channel is also attractive but mild, and corresponds to inter-orbital charge fluctuations. The SC channel is not at all attractive.
As the energy scale is lowered, the CDW channel is reduced because of charge screening. Around the scale $\Lambda\sim 1$, the SDW channel begins to be enhanced, establishing short-range spin exchanges. Concurrently, the CDW develops a cusp and begins to be enhanced at lower energy scales. This cusp follows from a level crossing from onsite to bond charge fluctuation modes (see below). The SC channel also begins to become attractive and enhanced.
As the energy scale decreases further, all channels are enhanced, and eventually the SC diverges first, indicating a normal state instability toward a primary SC order.

\begin{figure}
\includegraphics[width=\linewidth]{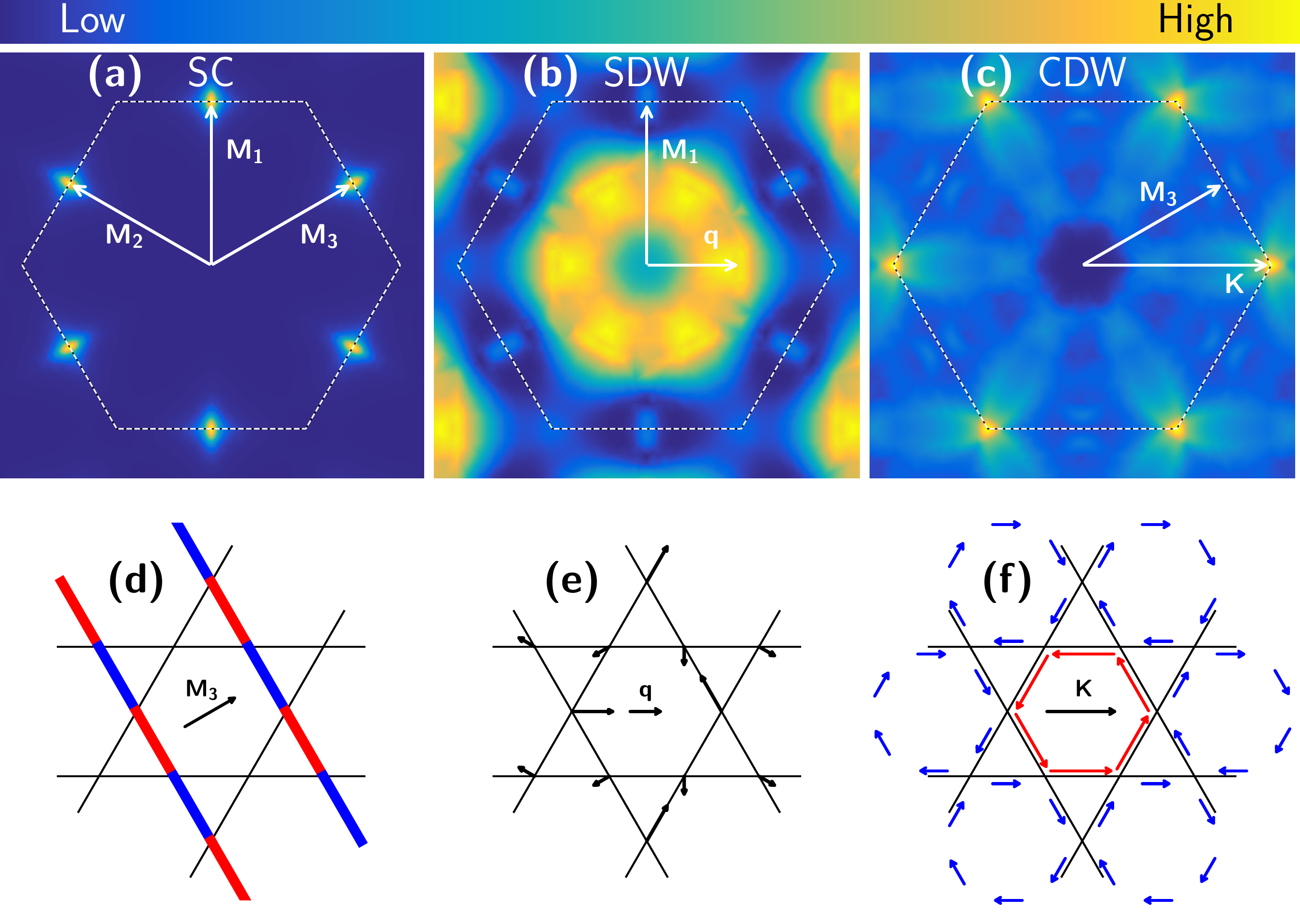}
\caption{The leading effective interactions at the divergent energy scale in the SC (a), SDW (b) and CDW (c) channels. (They are normalized by the respective maximum for a better view.) The white hexagon is the Brillouin zone. The leading eigen modes for the $y$-orbital at the peak momentum, represented in real space, are shown in (d) for PDW, (e) for SDW, and (f) for CDW. In (d) the blue and red bonds denote spin-singlet pairing at $\0M_3$ with opposite sign, in (e) the arrows denote the non-collinear spins at $\0q_{sdw}\approx\0K/2$, and in (f) the arrows denote the directions and amplitudes of the bond current at $\0K$ with a $3\times3$ period.}
\label{fig:sq}
\end{figure}

\begin{figure*}
	\includegraphics[width=\linewidth]{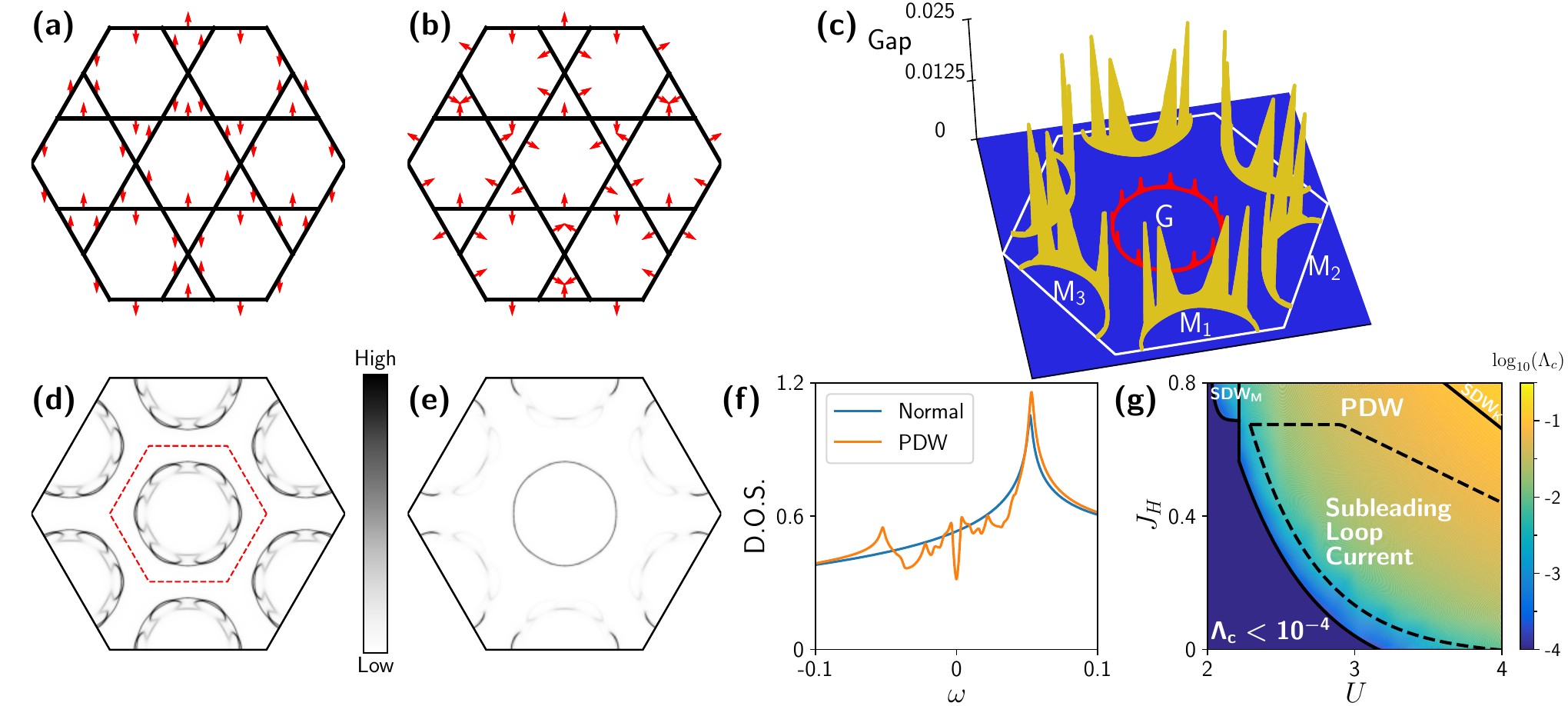}
	\caption{(a) Collinear and (b) non-collinear mean field PDW configurations. The arrows represent the phase and amplitude of the spin-singlet pairings on bonds. (c) The quasiparticle gap along the normal state Fermi pockets for a representative non-collinear chiral PDW. The gap on the $\0G$-pocket is multiplied by a factor of $3$ for better view. (d) Folded and (e) unfolded zero-energy spectral function in the momentum space. The red dashed hexagon in (d) is the folded Brillouin zone. (f) Comparison of the density of states in the PDW (orange line) and normal (blue line) states.	(g) FRG phase diagram in the $(U,J_H)$ plane, where the color scale encodes the critical energy scale, and the black dashed line encloses a regime with subleading loop-current instability within the PDW phase. }
	\label{fig:pdw}
\end{figure*}

Fig.~\ref{fig:flow}(b) shows that the SC correlation is intertwined with the correlations in the SDW and CDW channels under the FRG flow.
At the divergence scale, the SC is strongest, the SDW channel is the next strong, and the CDW channel is the weakest. To reveal the structure of the growing correlations, we plot the leading eigenvalue of the effective interactions as a function of momentum in Fig.~\ref{fig:sq}(a-c). The SC channel (a) shows clear and sharp peaks at the $\0M$ points, indicating the emerging SC order is indeed a primary PDW with CMM ${\0Q_{pdw}}= {\0M}_{1,2,3}$, consistent with the peak structure due to bond pairs in the pairing susceptibility $\chi_{pp}$ in Fig.~\ref{fig:flow}(a).
The SDW channel (b) shows broad and strong peaks close to $\0K/2$ and slightly weaker peaks at the $\0M$ points. These peak positions are consistent with the PH susceptibility $\chi_{ph}$ in Fig.~\ref{fig:flow}(a).
The CDW channel (c) is dominated by bright peaks at the $\0K$ points in contrast to the broad peak structures in the bare PH susceptibility $\chi_{ph}$ due to the correlation effects.

The leading eigen scattering modes associated with the strongest peaks in Fig.~\ref{fig:sq}(a-c) are shown in Fig.~\ref{fig:sq}(d-f). The numerical details for such modes are presented in Sec.II of SM \cite{sm}. Since the $x$ orbital content turns out to be much weaker than that of the $y$ orbital, only the $y$ orbital component is presented for clarity.
In Fig.~\ref{fig:sq}(d) for the SC channel, the eigen mode at ${\0M}_3$ is presented. The red and blue (connecting sublattices 1 and 2) correspond to spin-singlet pairing on bonds with opposite signs. The pairing also changes sign upon unit cell translation according to the plane wave $e^{i{\0M}_3\cdot {\0R}}$, where ${\0R}$ denotes the unit cell position. Note there are three degenerate PDW modes at the three $\0M$'s, which are simply related by $C_6$ rotations. In Fig.~\ref{fig:sq}(e) for the SDW channel, the eigen mode is a non-collinear SDW represented by the arrows on the lattice sites, which carries the momentum ${\0q}_{sdw}\approx{\0K}/2$ as highlighted in Fig.~\ref{fig:sq}(b). Finally, in Fig.~\ref{fig:sq}(f) for the CDW channel, the leading eigen mode has a $3\times3$ bond current, indicated by the length and direction of the arrows. We should note here that while Fig.~\ref{fig:sq}(d)
represents the pattern of the PDW order, Fig.~\ref{fig:sq}(e,f) can only be taken as representing the pattern of correlations in the SDW and CDW channels that are not yet divergent.

The present model exhibits novel intertwined correlations under the FRG that are beyond common expectations. First, the PDW pairing field $\Delta_{\0M_i}$ is expected to induce a secondary CDW field $\rho_{\0M_i+\0M_j}\sim \Delta_{\0M_i} \Delta_{-\0M_j}^*$, either at the reciprocal lattice wavevectors $\0G=2\0M_i$ ($i=j$) or at another $2\times2$ momentum $\0M_k=\0M_i+\0M_j$ ($i\neq j$). The FRG results in Fig.~\ref{fig:sq}(c) clearly shows that these peaks are absent, possibly because the induced CDW correlations are primarily onsite charge density modulations and are suppressed by the onsite Coulomb repulsion $U$. Instead, the induced CDW mode peaked at the $\0K$ points exhibits intriguing persistent loop current correlations as shown in Fig.~\ref{fig:sq}(c,f). Since the SDW mode has momentum $\0q_{sdw}\approx\0K/2$, it is highly suggestive that the CDW mode is instead induced by the SDW correlations even though it is subleading to the paring correlations and has not reached divergence. The spatial pattern in Fig.~\ref{fig:sq}(e) is described by $\textbf{S}(\0r)=\textbf{S}_{\0q_{sdw}}e^{i\0q_{sdw}\cdot\0r}+c.c.$. The induced CDW correlations is $\rho_{\0q_{cdw}}\sim \textbf{S}_{\0q_{sdw}}\cdot \textbf{S}_{\0q_{sdw}}$ with $\0q_{cdw}=2\0q_{sdw}=\0K$. Note that if the spins were collinear, then $\textbf{S}_{\0q_{sdw}}$ would be real and the induced CDW would be real. However, since the SDW at $\0q_{sdw}$ is non-collinear in Fig.~\ref{fig:sq}(e), which implies that $\textbf{S}_{\0q_{sdw}}$ is complex. Under the condition that the non-collinear spin configuration is not a circular spiral, $\textbf{S}_{\0q_{sdw}}\cdot \textbf{S}_{\0q_{sdw}}$ is complex, leading to a complex bond CDW at $2\0q_{sdw}=\0K$ with loop current as produced by the FRG in Fig.~\ref{fig:sq}(f). Interestingly, we note that there is a weaker peak at the $\0M$ points in the SDW channel in Fig.~\ref{fig:sq}(b). This offers an alternative scenario of multi-Q SDW correlations that can induce a CDW pattern at $\0K$, i.e. $\rho_\0K\sim \textbf{S}_{\0M_3} \cdot \textbf{S}_{\0q_{sdw}}$ as depicted in Fig.~\ref{fig:sq}(c) and under $C_6$ rotations, which is generically complex with bond currents.

Once the correlations in the SC channel diverges in the FRG, it is legitimate to use the PDW eigen modes to construct an effective mean field theory, see Sec. II of SM \cite{sm} for details. This helps to see how the three modes, now the multiple components of the SC order parameter, combine to form the $2\times2$ triple-Q PDW ordered state. We find that all components participate equally, but the relative phase between them can be either in-phase (collinear) or at $120^o$ (non-collinear), as indicated by the arrows in Fig.~\ref{fig:pdw}(a) and \ref{fig:pdw}(b), respectively. In both cases the pairing order parameters average to zero, while the non-collinear state additionally breaks time-reversal symmetry and is a chiral PDW state.
Note that the three-component PDW states are commensurate with the lattice and are described by $U(1)\times Z_2\times Z_2$ symmetry, where the continuous $U(1)$ is the overall phase of the PDW superconductor, and the discrete $Z_2$ symmetries arise from the translation by the lattice vectors in the $2\times2$ unit cell that shifts the phase by $\pi$, corresponding to a sign
change in one of the order parameter components. The discrete symmetry leads to four types of PDW domains and the proliferation of the fluctuating $Z_2$ domain walls can destroy the long-range PDW order and produce interesting vestigial cluster pairing states with higher-charge flux quantization \cite{Zhou2022,Varma2023,Ge2024,Wang_A_2025}.

We further discuss the single-particle properties of the chiral PDW state, see Sec. III of SM \cite{sm} for technical details. The quasiparticle gaps (on the occupied side) around $\0G$ and $\0M$ pockets are shown in Fig.~\ref{fig:pdw}(c). Both are very anisotropic. The gap on the $\0G$-pocket is much smaller, and can be related to the fact that the inter-pocket pairing between $\0G$ and $\0M$ lacks quasi-nesting, and suffers from weak sublattice selectivity. The zero-energy spectral function in the folded Brillouin zone is shown in Fig.~\ref{fig:pdw}(d). All pockets are folded to embrace the zone center, and in this way it is clear that the spectral weight is suppressed most strongly where the folded pockets intersect (hence the gap is larger). Fig.~\ref{fig:pdw}(e) shows the zero-energy spectral function in the unfolded zone, which is directly relevant to angle-resolved photoemission spectroscopy. Here the $\0G$-pocket is most weakly suppressed, and the $\0M$-pockets are also weakly suppressed where the gap is small. The weakly suppressed parts of the pockets would be indistinguishable with the ``residual'' Fermi surface as a generic feature of PDW superconductors, reminiscent of KV$_3$Sb$_5$ and CsV$_3$Sb$_5$ \cite{Deng2024,Mine2025}.
In Fig.~\ref{fig:pdw}(f), we plot the density of states (DOS) in the chiral PDW state (orange line) in comparison to the normal state (blue line). The low energy DOS is only partially suppressed, and there are some spikes resulting from the anisotropic gap as shown in Fig.~\ref{fig:pdw}(c).

Finally, by systematic SM-FRG calculations, we obtain the phase diagram Fig.~\ref{fig:pdw}(g) in the $(U, J_H)$ parameter space. The PDW state is obtained as the ground state in a large regime of moderate interactions. Inside the PDW phase, the dashed line encloses a regime with subleading loop-current correlations. At even large $U$ and $J_H$, the ordered state becomes a primary SDW with ordering wavevector at $\0K$. No instabilities are found for weaker interactions (at least for $\Lambda>10^{-4}$), indicating a stable paramagnetic phase. The threshold boundary is reasonable since in the present model, the PDW/SDW/CDW is free from infrared divergence in the susceptibility. \footnote{By the Kohn-Luttinger anomaly argument, a zero-momentum SC state at infinitesimal energy scales in certain high angular momentum channels are possible, but it is unrelated to the current discussions here.} We also find that the primary PDW state is robust against variations in the band filling, and hence variations in the size of Fermi pockets, see Sec. IV of SM \cite{sm}.

\emph{Summary}.
We proposed a two-orbital Hubbard model on the kagome lattice and demonstrated by unbiased FRG that the long-sought primary PDW state is realized under fairly realistic physical conditions, including the local Coulomb interactions as well as the finite-sized Fermi pockets. The multiple Fermi pockets with well defined orbital and sublattice selectivity suppresses even the quasi-nesting for PH scattering, opening the door for primary PDW order. While the central $\0G$-pocket is barely gapped by the PDW, we find its presence is necessary to remove the otherwise tendency toward ferromagnetic order \cite{Wang2013}, which would be unfavorable for spin-singlet pairing. The finite CMM pairing for the primary PDW order is found to arise from spin fluctuations and is strongly intertwined with both spin and charge (current) correlations. We propose that the model can be realized in certain $p$- and $d$-orbital kagome materials such as CsCr$_3$Sb$_5$ \cite{CsCr3Sb5}, which contain dominant Fermi pockets that closely resemble those of the current two-orbital model.
Our findings reveal a new path toward novel quantum matter via integrated sublattice and orbital degrees of freedom in correlated quantum materials.

\emph{Acknowledgments}.
H.Y.L., D.W. and Q.H.W. are supported by National Key R\&D Program of China (Grants No. 2022YFA1403201, No. 2024YFA1408104), National Natural Science Foundation of China (Grants No. 12374147, No. 12274205, No. 92365203).
Z.W. is supported by the U.S. Department of Energy, Basic Energy Sciences Grant DE-FG02-99ER45747.

\bibliography{pdw}

\vspace{20pt}

\begin{center}{\Large \textbf{
Supplemental Materials on ``Genuine pair density wave order on the kagome lattice''
}}\end{center}

\setcounter{figure}{0} 
\renewcommand{\thefigure}{S\arabic{figure}} 
\renewcommand{\thetable}{S\arabic{table}} 
\renewcommand{\theequation}{S\arabic{equation}}

In this Supplemental Materials, we explain some technical details and present additional  data referred to in the main text. (i) First, we explain the technical details of the singular-mode functional renormalization group (SM-FRG) applied in the main text. This includes how the interaction vertex is decomposed into scattering matrices in the fermion bilinear basis, how the FRG flow equations are defined in terms of the scattering matrices, and how the leading scattering eigenmodes in the superconductivity (SC), spin density wave (SDW) and charge density wave (CDW) channels are defined. (ii) We tabulate the leading scattering modes in the SC, SDW and CDW channels discussed in the main text for concreteness. (iii) Next, we explain how the divergent scattering modes from FRG are used to construct an effective mean field theory, by which we can calculate the single-particle properties of the pair density wave (PDW) state. (iv) Finally, we present the phase diagram of the model at various filling levels to show that the primary PDW state we discussed in the main text is quite generic.

\section{Singular-mode FRG}
As one of the powerful methods to study correlated electronic systems, the functional renormalization group (FRG) has been well explained in many documents \cite{Berges_PR_2002,Metzner_RMP_2012,Dupuis_PR_2021,Kopietz__2010}. In this Supplemental Materials, we mainly focus on one of its realizations, called singular-mode FRG (SM-FRG).

\begin{figure}
	\includegraphics[width=0.9\linewidth]{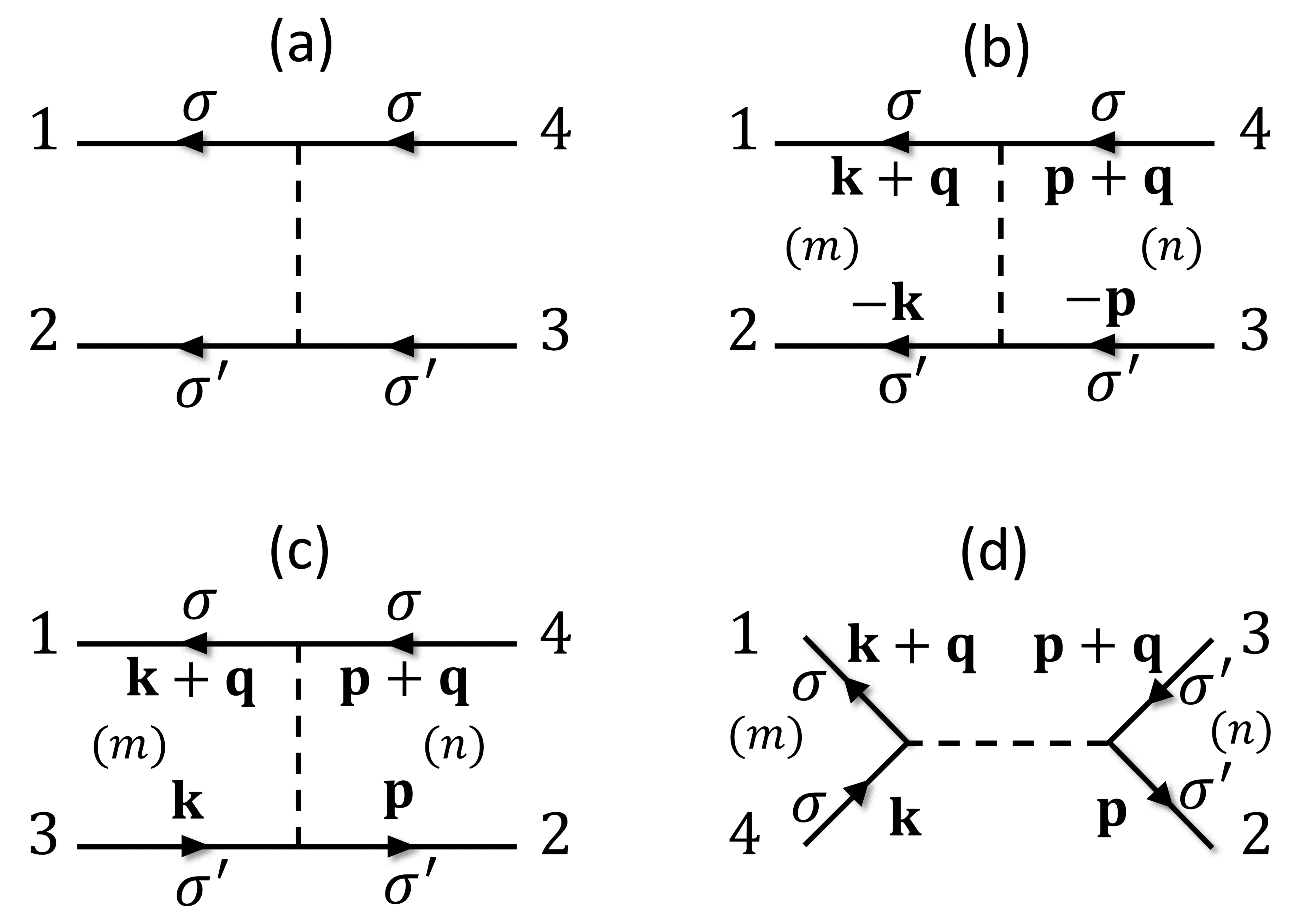}
	\caption{A generic 4-point 1PI vertex (a) can be rearranged into the pairing (P), crossing (C) and direct (D) channels as shown in (b)-(d), respectively. The momentum $\mathbf{k,q,p}$ are explicitly shown for clarity. The spins ($\sigma$ and $\tau$) are conserved during fermion propagation in the spin-SU(2) symmetric case. The labels $m$ and $n$ denote fermion bilinears.}
	\label{fig:vertex}
\end{figure}

\subsection{Decomposition of interaction vertex in the fermion bilinear basis}
Consider the 4-point one-particle irreducible (1PI) vertices $\Gamma_{1234}$ appearing in the effective interaction
\begin{equation} H_{\Gamma}=\frac{1}{2}\sum_{1,2,3,4}\psi^\dagger_1\psi^\dagger_2\Gamma_{1234}\psi_3\psi_4,\end{equation}
where $\psi$ is the fermion field with the subscript $1,2,3,4$ denoting the one-particle degrees of freedom (such as frequency, momentum, orbital, sublattice, spin, etc.). In the SU(2) symmetric case under concern, the spins for 1 and 4 are the same, and similarly for 2 and 3.
We define fermion bilinears in the three Mandelstam channels as,
\begin{align}
\alpha_{12}^{\dag}&=\psi_1^\dag\psi_2^\dag \ \text{ (pairing)}, \\ \beta_{13}^{\dag}&=\psi_1^\dag\psi_3\  \text{ (crossing)}, \\ \gamma_{14}^{\dag}&=\psi_1^\dag\psi_4 \ \text{ (direct)}.
\end{align}
Then the general 1PI vertex can be rewritten as scattering matrices $P$, $C$ and $D$ in the three channels as
\begin{align}
H_\Gamma&=\frac12 \sum_{12,43}\alpha_{12}^{\dag} ~P_{12;43}~\alpha_{43} \nonumber\\
&= -\frac12 \sum_{13,42}\beta_{13}^{\dag} ~C_{13;42}~ \beta_{42} \nonumber\\
&= \frac12 \sum_{14,32}\gamma_{14}^{\dag} ~D_{14;32} ~\gamma_{32}, \label{eq:rewind}
\end{align}
as illustrated in Fig.~\ref*{fig:vertex}(b)-(d).
The momentum of the fermion bilinear is
$\0q = \0k_1+\0k_2$, $\0k_1-\0k_3$, $\0k_1-\0k_4$, in the $P$, $C$ and $D$ channels.
If the subscripts $1,2,3,4$ run over all sites, the three scattering matrices $P$, $C$ and $D$ are all equivalent to $\Gamma$, i.e. $\Gamma_{1234}=P_{12;43} =C_{13;42}=D_{14;32}$.
But in practical calculations, the fermion bilinears must be truncated. On physical grounds, the important bilinears are those that join the singular scattering modes, and such eigenmodes determine the emerging order parameter. Since order parameters are composed of short-ranged bilinears, only such bilinears are important. These include onsite and on-bond pairing in the pairing channel, and onsite and on-bond particle-hole density in the C and D channels. Note the truncation is applied only for the internal distance between the two fermions in a fermion bilinear, while the setback distance between two bilienars can be arbitrary. In this way, the thermodynamic limit can be reached for the propagation of fermion bilinears. The truncation range of bonds is satisfactory if longer bonds do not lead to qualitative change of the results. This is guaranteed since the decomposition is exact in all channels in the asymptotic limit where a complete set of the fermion bilinears is used. The FRG based on the decomposition of the interaction vertices into scattering matrices in the truncated fermion bilinear basis, which are sufficient to capture the most singular scattering modes, is called the singular-mode FRG (SM-FRG) \cite{Wang2012,Xiang_PRB_2012,Wang2013}.
\begin{figure}
	\includegraphics[width=0.9\linewidth]{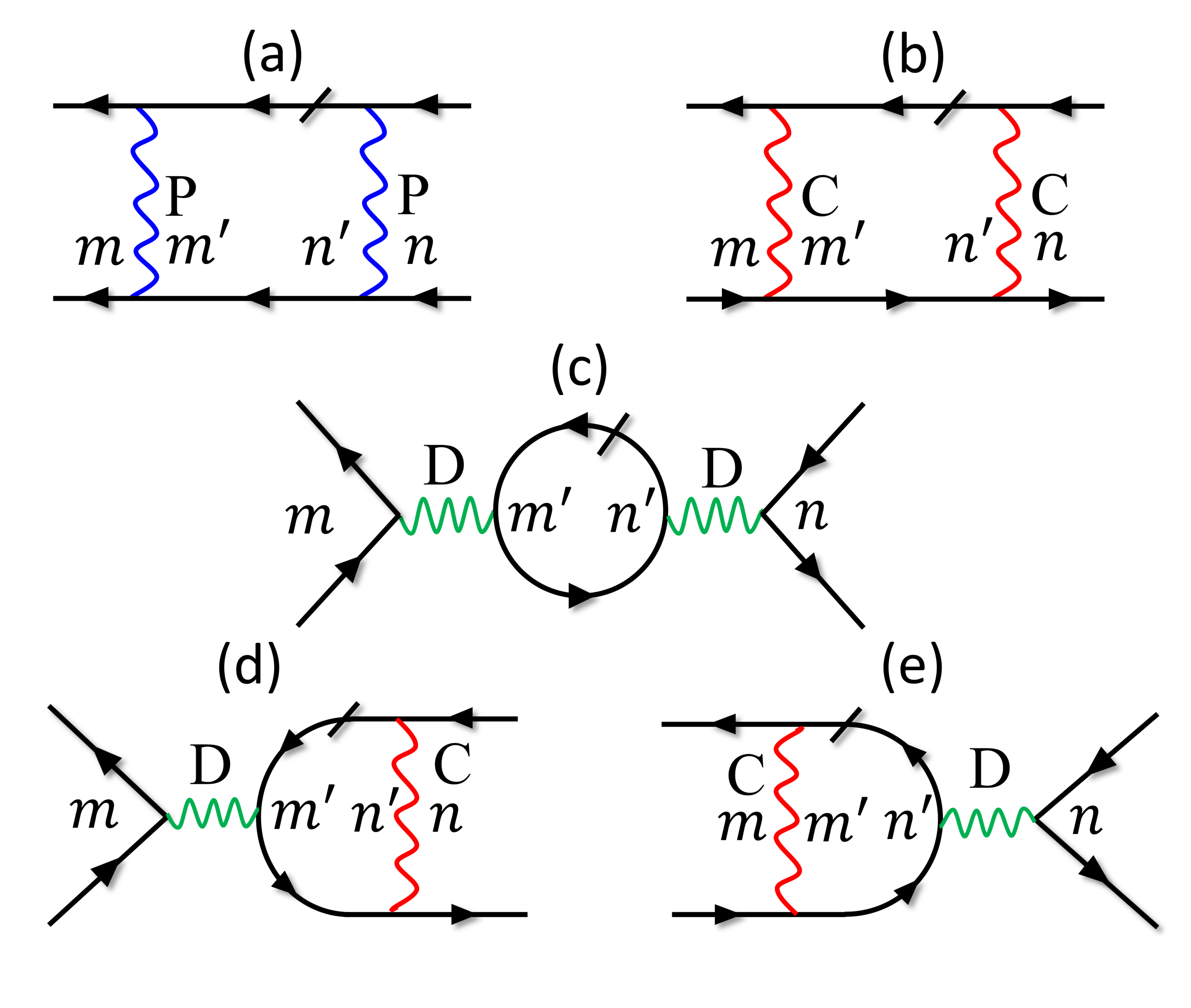}
	\caption{ One-loop contributions to $\partial\Gamma_{1234}/\partial\Lambda$. The wavy lines denote the truncated $P$ (blue), $C$ (red) and $D$ (green) from $\Gamma$. The slash in each diagram denotes the single-scale propagator and is put on either one of the fermion lines within the loop. The symbols $m$ and $n$ denote the fermion bilinear in the respective channels, and the inner indices are summed implicitly.}
	\label{fig:1loop}
\end{figure}

\subsection{Flow equations}
Starting from $\Lambda=\infty$ where the 1PI vertices $P$, $C$ and $D$ are given by the bare interactions, $\Gamma_{1234}$ flows as
\begin{align} \label{eq:flow}
\frac{\partial\Gamma_{1234}}{\partial\Lambda} &= [P\chi_{pp}P]_{12;43}+[C\chi_{ph}C]_{13;42}\nonumber\\ &+ [D\chi_{ph}C+C\chi_{ph}D-2D\chi_{ph}D]_{14;32} ,
\end{align}
see Fig.~\ref*{fig:1loop} for illustration.
The products within the square brackets imply matrix convolutions, and $\chi_{pp}$ and $\chi_{ph}$ are single-scale (at $\Lambda$) particle-particle and particle-hole susceptibilities given by, in real space,
\begin{align}
[\chi_{pp}]_{ab;cd} &= \frac{1}{2\pi}\left[ G_{ac}(i\Lambda)G_{bd}(-i\Lambda)+
(\Lambda \to -\Lambda) \right] ,\\
[\chi_{ph}]_{ab;cd} &= \frac{1}{2\pi}\left[ G_{ac}(i\Lambda)G_{db}(i\Lambda)+
(\Lambda \to -\Lambda) \right],
\end{align}
where $a,b,c,d$ are dummy fermion indices (that enter the fermion bilinear labels), and $G_{ab}(i\Lambda)$ is the normal state Matsubara Green's function. (The expression in the momentum space is slightly more complicated but is otherwise straightforward, and is used in real calculations.)
As usual \cite{Metzner_RMP_2012}, we neglect self-energy correction which could be absorbed in the band dispersion, and we also neglect the sixth and higher order vertices, which are RG-irrelevant. The frequency dependence of the 4-point vertices is also RG-irrelevant and ignored. In this spirit, all external legs are set at zero frequency.
The functional flow equation is solved by numerical integration over $\Lambda$.
Note that after $\Gamma_{1234}$ is updated after an integration step, it is rewinded as $P$, $C$ and $D$ according to $\Gamma_{1234}=P_{12;43} =C_{13;42}=D_{14;32}$, subject to truncation of the fermion bilinears.

The SM-FRG can treat interactions in all channels on equal footing. In fact, if we ignore the channel overlaps, the flow equation reduces to the ladder equation in the $P$-channel, and the random-phase-approximation in the $C$- and $D$-channels. The SM-FRG combines the three channels coherently. The advantage of SM-FRG also includes: (i) The momentum conservation law is respected exactly; (ii) The orbital-sublattice content of the Bloch states are kept exactly, since the calculation is performed in the orbital-sublattice basis. This point is crucial if the orbital-sublattice content varies quickly in the momentum space, as in model of the main text.

\subsection{Effective interactions and singular eigenmodes in SC/SDW/CDW channels}
The scattering matrices in the superconductivity (SC), spin density wave (SDW) and charge density wave (CDW) channels can be shown to be related to $P$, $C$ and $D$ as follows,
\begin{align}
V^{\rm SC}=P,\ \ V^{\rm SDW}=-C,\ \ V^{\rm CDW}=2D-C.
\end{align}
In a given channel, the matrix can be decomposed by singular value decomposition (SVD), in momentum space,
\begin{align}
V_{mn}(\0q)=\sum_{\alpha} \phi_{\alpha m}(\0q) S_{\alpha}(\0q) \phi^*_{\alpha n}(\0q),
\end{align}
where $m,n$ label the fermion bilinear, $S_\alpha$ and $\phi_{\alpha n}$ are eigenvalue and eigenvector for the $\alpha$-th singular mode. In each channel, the number of eigenmodes for each momentum $\0q$ equals the number of fermion bilinears used to expand the interactions.

During the SM-FRG flow, we monitor the leading (most negative) eigenvalue, which we abbreviate as $S$ in each channel.
As the energy scale $\Lambda$ reduces, the first divergence of $S$ indicates a tendency towards an instability with order parameter described by the associated eigenmode $\phi(\0Q)$, where $\0Q$ is the corresponding collective momentum. In this case, one can drop the nonsingular components to write the renormalized interaction as,
\begin{equation}
H_\Gamma \sim \frac{S}{N}O^\dagger O + \cdots,
\end{equation}
where $N$ is the number of unitcells, $O$ is the mode operator that is a combination of the fermion bilinears (see below), and the dots represent symmetry related terms. For example, if the SC channel diverges first, we have
\begin{eqnarray}
O_{SC}^\dagger & =& \sum_n \phi_n(\0Q) \alpha_n^\dagger(\0Q) \nonumber\\ &\to & \sum_{\0k,n=(a,b,\bm{\delta})}\psi_{\0k+\0Q, a}^\dagger \phi_n(\0Q)  e^{i\0k\cdot\bm{\delta}}\psi_{-\0k,b}^\dagger\nonumber\\
&\to& \sum_{\0r,n=(a,b,\bm{\delta})} e^{i\0Q\cdot \0r}\psi_a^\dagger(\0r)\phi_n(\0Q)\psi_b^\dagger(\0r+\bm{\delta}),\label{eq:pair}\end{eqnarray}
where $n$ labels a fermion bilinear, $a$ and $b$ denote the orbital (and sublattice). The second line is the form in momentum space, and the last line in real space, with $\phi_{n=(a,b,\bm{\delta})}(\0Q)$ acting as the element of the real-space pairing matrix on the bond $\bm{\delta}$ radiating from orbital $a$ at position $\0r$ to $b$ at $\0r+\bm{\delta}$.
The spin indices do not have to be specified, as the symmetry of the gap function under inversion automatically determines whether the pair is in the singlet or triplet state.

Similarly, if the SDW channel diverges first, we obtain the mode operator
\begin{eqnarray}
O_{SDW}^\dagger  &=& \sum_n \phi_n(\0Q)\beta_n^\dagger(\0Q)\nonumber\\ &\to&
\sum_{\0k,n=(a,b,\bm{\delta})}\psi_{\0k+\0Q,a,\uparrow}^\dagger \phi_n(\0Q) e^{i\0k\cdot\bm{\delta}}\psi_{\0k,b,\downarrow} \nonumber\\
&\to& \sum_{\0r,n=(a,b,\bm{\delta})}e^{i\0Q\cdot\0 r}\psi_{a\uparrow}^\dagger(\0r)\phi_n(\0Q)\psi_{b\downarrow}(\0r+\bm{\delta}) ,\nn\\
\end{eqnarray}
where we assign the spin order in the transverse direction.
Finally, if the CDW channel diverges first, we obtain the mode operator
\begin{eqnarray}
O_{CDW}^\dagger &=& \sum_n \phi_n(\0Q)\gamma_n^\dagger(\0Q) \\ &\to&
\sum_{\0k,\sigma,n=(a,b,\bm{\delta})}\psi_{\0k+\0Q,a,\sigma}^\dagger \phi_n(\0Q) e^{i\0k\cdot\bm{\delta}}\psi_{\0k,b,\sigma}\nonumber\\
&\to&\sum_{\0r,\sigma,n=(a,b,\bm{\delta})}e^{i\0Q\cdot\0r}\psi_{a\sigma}^\dagger(\0r)\phi_n(\0Q)\psi_{b\sigma}(\0r+\bm{\delta}). \nn\\
\end{eqnarray}
Note that $H_{SDW/CDW}$ can capture both onsite and on-bond density waves, since the fermion bilinears contain both cases of $\bm{\delta}=\00$ and $\bm{\delta}\neq \00$.

\phantomsection

\section{Representative leading eigenmodes discussed in the main text}
Here we present the explicit forms of the leading eigenmodes discussed in the main text. The eigenmode is characterized by
\begin{align} \label{eq:f}
f^{ab}_{\bm{\delta}}=e^{i\0Q\cdot \0r_a}\,\phi_{(a,b,\bm{\delta})}(\mathbf{Q}),
\end{align}
where $a$ and $b$ are the combined indices denoting (orbital, sublattice), and $\bm{\delta}$ denotes the bond vector from $a$ to $b$. In this work on the kagome lattice, we take the three sublattices in the primitive cell to be located at
\begin{align}
\0r_1=(0,0),\qquad
\0r_2=\left(-\frac14,\frac{\sqrt3}{4}\right),\qquad
\0r_3=\left(\frac14,\frac{\sqrt3}{4}\right).
\end{align}
In the strongest SC channel at $\0Q=\0M_3$, the main elements of this eigenmode are listed in Table~\ref{tab:S1}. We note that apart from the bond-pair component, the last two lines are from onsite intra-orbital pairs. They are antiphase on the two orbitals, are related to and enhanced by the Hund's coupling.

Similarly, the dominant elements of the leading SDW eigenmode at $\0Q=\0K/2$ are listed in Table~\ref{tab:S2}. In this case, the local bilinears dominate, indicating predominantly site-local spin polarization.

Finally, the dominant elements of the leading CDW eigenmode at $\0Q=\0K$ are listed in Table~\ref{tab:S3}. Here the bond bilinears dominate, and the odd parity under site exchange indicate the eigenmode represents bond current.

\begin{table}
\caption{Dominant elements of the leading pairing eigenmode at $\0Q=\0M_3$, as characterized by $f^{ab}_{\bm{\delta}}$ with $a|b$=(orbital, sublattice) and $\bm{\delta}$ the bond vector from $a$ to $b$.}
\label{tab:S1}

\begin{ruledtabular}
\begin{tabular}{cccc}
$a$ & $b$ & $\bm{\delta}$ & $f_{\bm{\delta}}^{ab}$ \\
\hline
$(y,1)$ & $(y,2)$ & $(-1/4,\,\sqrt{3}/4)$  & $0.47318$ \\
$(y,2)$ & $(y,1)$ & $(1/4,\,-\sqrt{3}/4)$  & $0.47318$ \\
&&&\\
$(y,1)$ & $(y,2)$ & $(1/4,\,-\sqrt{3}/4)$  & $-0.47318$ \\
$(y,2)$ & $(y,1)$ & $(-1/4,\,\sqrt{3}/4)$  & $-0.47318$ \\
&&&\\
$(y,1)$ & $(x,2)$ & $(-1/4,\,\sqrt{3}/4)$  & $-0.09643$ \\
$(x,2)$ & $(y,1)$ & $(1/4,\,-\sqrt{3}/4)$  & $-0.09643$ \\
&&&\\
$(x,1)$ & $(y,2)$ & $(-1/4,\,\sqrt{3}/4)$  & $0.09643$ \\
$(y,2)$ & $(x,1)$ & $(1/4,\,-\sqrt{3}/4)$  & $0.09643$ \\
&&&\\
$(y,1)$ & $(x,2)$ & $(1/4,\,-\sqrt{3}/4)$  & $0.09643$ \\
$(x,1)$ & $(y,2)$ & $(1/4,\,-\sqrt{3}/4)$  & $-0.09643$ \\

$(x,2)$ & $(y,1)$ & $(-1/4,\,\sqrt{3}/4)$  & $0.09643$ \\
$(y,2)$ & $(x,1)$ & $(-1/4,\,\sqrt{3}/4)$  & $-0.09643$ \\
&&&\\
$(y,3)$ & $(y,3)$ & $(0,\, 0)$  & $-0.08643$ \\
$(x,3)$ & $(x,3)$ & $(0,\, 0)$  & $0.04508$ \\
\end{tabular}
\end{ruledtabular}
\end{table}

\begin{table}
\caption{Similar to Table~\ref{tab:S1} but for the leading SDW eigenmode at $\0Q=\0K/2$.}
\label{tab:S2}
\begin{ruledtabular}
\begin{tabular}{cccc}
$a$ & $b$ & $\bm{\delta}$ & $f_{\bm{\delta}}^{ab}$ \\
\hline
$(y,1)$ & $(y,1)$ & $(0,\,0)$ & $0.74393$ \\
$(y,2)$ & $(y,2)$ & $(0,\,0)$ & $-0.36258\, e^{-i \pi/6}$ \\
$(y,3)$ & $(y,3)$ & $(0,\,0)$ & $-0.36258\, e^{i \pi/6}$ \\
$(x,1)$ & $(x,1)$ & $(0,\,0)$ & $0.30087$ \\
\end{tabular}
\end{ruledtabular}
\end{table}

\begin{table}
\caption{Similar to Table~\ref{tab:S1} but for the leading CDW eigenmode at $\0Q=\0K$.}
\label{tab:S3}
\begin{ruledtabular}
\begin{tabular}{cccc}
$a$ & $b$ & $\bm{\delta}$ & $f_{\bm{\delta}}^{ab}$ \\
\hline
$(y,1)$ & $(y,2)$ & $(-1/4,\,\sqrt{3}/4)$  & $-0.25036$ \\
$(y,2)$ & $(y,1)$ & $(1/4,\,-\sqrt{3}/4)$  & $0.25036$ \\
$(y,1)$ & $(y,3)$ & $(1/4,\,\sqrt{3}/4)$  & $-0.25036\, e^{i \pi/3}$ \\
$(y,3)$ & $(y,1)$ & $(-1/4,\,-\sqrt{3}/4)$  & $0.25036\, e^{i \pi/3}$ \\
$(y,2)$ & $(y,3)$ & $(1/2,\,0)$  & $0.25036 \, e^{-i \pi/3}$ \\
$(y,3)$ & $(y,2)$ & $(-1/2,\,0)$  & $-0.25036 \, e^{-i \pi/3}$ \\
&&&\\
$(y,1)$ & $(y,2)$ & $(1/4,\,-\sqrt{3}/4)$  & $0.25036\, e^{i \pi/3}$ \\
$(y,1)$ & $(y,3)$ & $(-1/4,\,-\sqrt{3}/4)$  & $0.25036$ \\
$(y,2)$ & $(y,1)$ & $(-1/4,\,\sqrt{3}/4)$  & $-0.25036\, e^{-i \pi/3}$ \\
$(y,2)$ & $(y,3)$ & $(-1/2,\,0)$  & $-0.25036$ \\
$(y,3)$ & $(y,1)$ & $(1/4,\,\sqrt{3}/4)$  & $0.25036\, e^{-i \pi/3}$ \\
$(y,3)$ & $(y,2)$ & $(1/2,\,0)$  & $-0.25036\, e^{i \pi/3}$ \\
\end{tabular}
\end{ruledtabular}
\end{table}

\section{FRG-derived mean field theory for the PDW state}
According to FRG, there are three degenerate singular modes in the SC channel. This can be modeled by an effective mean field theory. The mean field Hamiltonian can be written as $H_{MF}=H_0+H_P$, where $H_0$ is the normal part, and $H_P$ is the pairing part, in the spirit of the above discussion,
\begin{equation}
H_P=-\sum_{\0Q} \Delta_\0Q B_\0Q^\dagger + { h.c.} \ . \end{equation}
Here $\Delta_\0Q$ has the dimension of energy acting as the order parameter for the PDW momentum $\0Q=\0M_{1,2,3}$, and $B_\0Q^\dagger$ is of the form in Eq.~\ref{eq:pair}. The self-consistent condition for the order parameter is
\begin{equation} \Delta_\0Q = -\frac{V}{N}\langle B_\0Q\rangle, \end{equation}
where $V>0$ is the effective pairing strength, which can be tuned to match the divergence scale $\Lambda_c$ in FRG, but may also be taken as a tuning parameter for qualitative purposes. In the subsequent calculations for the single-particle properties, we simply set $|\Delta_M|=0.02$ for illustration.

Since the PDW is 2$\times$2 periodic, it is more convenient to write the mean field Hamiltonian
in real space, and use 2$\times$2 as the supercell. The PDW then becomes formally a zero-momentum pairing in the reduced Brillouin zone. We use a 48-component spinor $\Psi=(\psi_\up,\psi_\downarrow^\dagger)$, where $\psi$ is a 24-component spinor describing electrons on all orbitals and lattice sites within a supercell.
Then the mean field Hamiltonian can be written as
\begin{equation} H_{MF}=\sum_{\0k\in RBZ} \Psi_\0k^\dagger h_\0k \Psi_\0k,\end{equation}
where $h_\0k$ is a 48x48 matrix, and $\0k$ is limited in the reduced Brillouin zone (RBZ), or folded zone.

\begin{figure}
	\includegraphics[width=\linewidth]{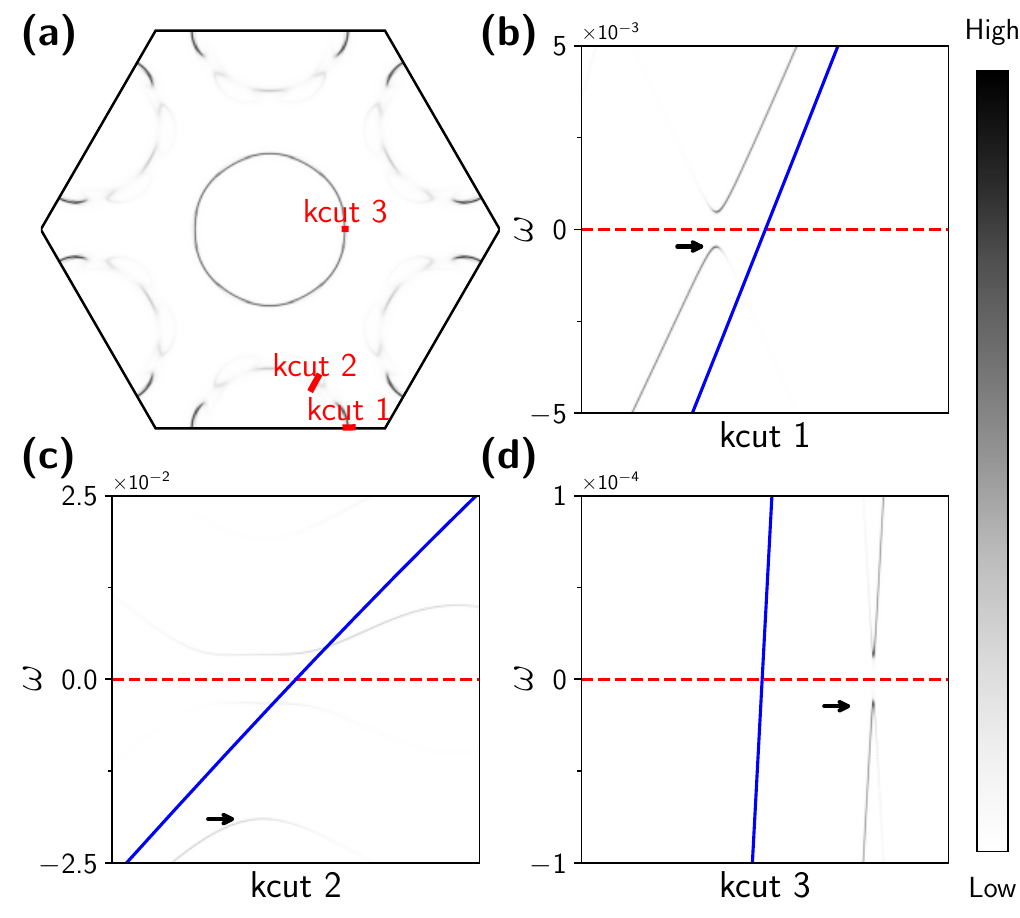}
	\caption{(a) Unfolded zero-energy spectral function, identical to {main text Fig.~4(e)}, with three representative line cuts indicated by the short red lines. Panels (b)–(d) show the unfolded electron spectral function (gray scale) along cuts 1–3. The blue line denotes the normal state dispersion, the red dashed line marks the Fermi level, and the black arrow indicates the extracted quasiparticle gap. To connect with ARPES, the gap is determined from the occupied ($\omega<0$) part of the spectrum.}
	\label{fig:akw}
\end{figure}

We can calculate the spectral function in the folded zone as
\begin{align}
A_{\rm folded}(\0k,\w)=-\frac{2}{\pi M}\text{Im} {\rm Tr}~G^{\up\up}(\0k,\w+i\eta),
\end{align}
where the factor of 2 accounts for spin degeneracy for singlet pairing, $M=24$ is the total number of single-particle degrees of freedom per spin (that is, 2 orbitals/site multiplied by 12 sites) within the supercell, $G^{\up\up}$ is the normal block of the retarded Green's function $G(\0k,\omega) = (\omega+i\eta -h_\0k)^{-1}$, where $\eta\to 0$ mimics the elastic scattering rate, and $\omega$ is the frequency.
However, experiment, such as angular-resolved photoemission spectroscopy (ARPES), actually measures the spectral function in the unfolded zone, which is given by
\begin{align}
A_{\rm unfolded}(\0k,\w)=-\frac{2}{\pi M}\text{Im} {\rm Tr}~ g^{\up\up}(\0k,\w+i\eta),
\end{align}
where $\0k$ is in the normal state BZ, and $g$ is the reduced Green's function defined as
\begin{equation}
g_{\0r,\0r'}(\0k,\omega) = \sum_{\0R,\0R'}G_{\0r+\0R,\0r'+\0R'},
\end{equation}
where $\0r$ and $\0r'$ refer to sites within the primitive cell, $\0R$ and $\0R'$ are translation vectors for primitive cells, so that both $\0R+\0r$ and $\0R'+\0r'$ exhausts sites within the supercell. (The orbital labels are left implicit for brevity). Note that this applies if and only if the Fourier transform of the fermion field is defined as
\begin{equation}
\psi(\0r) = \frac{1}{\sqrt{N}}\sum_\0k \psi_\0k e^{i\0k\cdot \0r},
\end{equation}
where $\0r$ is the actual position of the field. This convention is adopted in the entire process of our SM-FRG.

\begin{figure}
	\includegraphics[width=\linewidth]{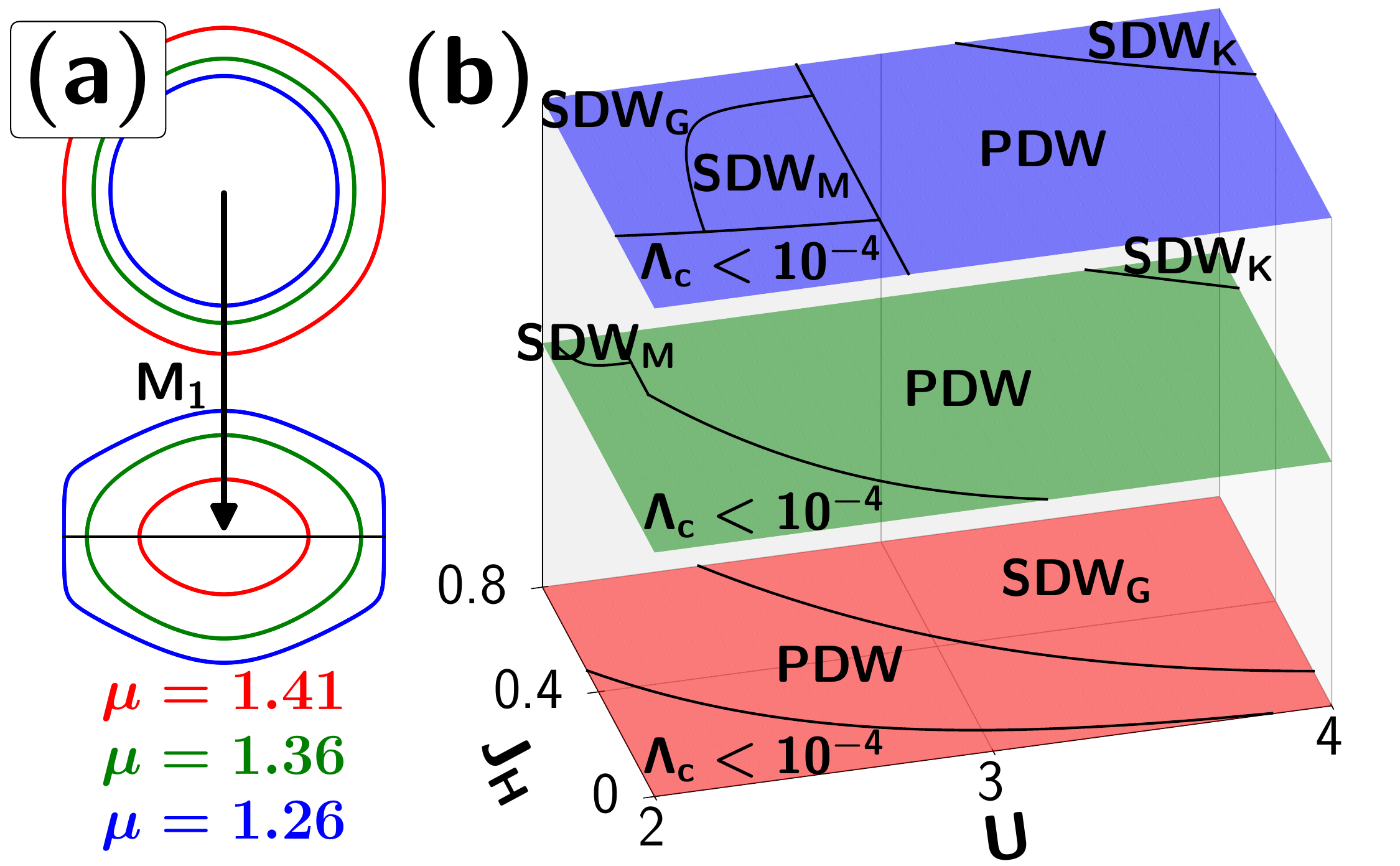}
	\caption{  (a) Fermi surfaces and (b) phase diagrams as a function of $U$ and $J_H$ for three chemical potentials. The subscript of SDW indicates the ordering wavevector, while the PDW always occurs at/near wavevector $\0M$.}
	\label{fig:phase}
\end{figure}

For uniform SC, the gap always opens at the normal state Fermi momentum $\0k_F$ exactly. But for the PDW, the quasiparticle energy minimum may shift away from $\0k_F$. The reason is the pairing block and the normal block in the Hamiltonian matrix $h_\0k$ do not commute, causing inter-band pairing.
To extract the quasiparticle gap, we evaluate the unfolded spectral function along line cuts (red thick lines) perpendicular to the normal-state Fermi surface shown in Fig.~\ref*{fig:akw}. The results (gray scale) are shown in Fig.~\ref*{fig:akw}(b-d), where we also show the normal state dispersion for comparison. Clearly, the PDW gap opening deviates slightly from the Fermi point, and the particle-hole symmetry is not present near the Fermi level. To be consistent with ARPES, we define the excitation gap as the quasiparticle peak below but closest to the Fermi level. This corresponds to the gap for particle-removing or hole excitation. We should point out, however, that the particle excitation (electron inserting) gap can be different to that for the hole excitation in the PDW state, as seen in the above line-cut plots. Both particle- and hole-excitation gaps can be revealed by STM.

\section{Phase diagrams at various filling levels}

Fig.~\ref*{fig:phase}(a) illustrates the Fermi pockets for three chemical potentials.  The relative pocket size changes. We performed FRG calculations for all of these filling levels. The corresponding phase diagrams are shown in Fig.~\ref*{fig:phase}(b). The regions with no instability up to a scale of $\Lambda =10^{-4}$ are also indicated. While the PDW is always at/near wavevector $\0M$, the SDW wavevector varies with both interactions and filling levels, which is indicated by the subscript. For example, SDW$_\0K$ means SDW at momentum $\0K$, and SDW$_\0G$ means SDW instability at momentum $\0G=\00$, which is actually unitcell-wise periodic.

\end{document}